\documentclass{aa}  

\usepackage{graphicx}
\usepackage{txfonts}
\newcommand{\dd}{\mathrm{d}}

\usepackage{amsmath}
\usepackage{amsfonts}
\usepackage{amssymb}
\usepackage{microtype}
\usepackage{graphics}
\usepackage{subfigure}
\usepackage{float}
\usepackage{color}
\usepackage{tikz}
\usetikzlibrary{decorations.pathreplacing,arrows,positioning,shapes.geometric,calc}
\usepackage{bm}
\usepackage{pgffor}
\usepackage{ulem}
\usepackage{url}
\usepackage{dsfont}

\begin{document}

  \title{Cosmological Constraints with Void Lensing}

  \subtitle{I. the Simulation-Based Inference Framework}

  \author{Chen Su\inst{1,2}\thanks{\email{suchen@shao.ac.cn}}
          \and
          Huanyuan Shan\inst{1,2}\thanks{\email{hyshan@shao.ac.cn}}
          \and
          Cheng Zhao\inst{3}\thanks{\email{czhao@tsinghua.edu.cn}}
          \and
          Wenshuo Xu\inst{3}
          \and
          Jiajun Zhang\inst{1,2}
          }

  \institute{
        Shanghai Astronomical Observatory (SHAO), Nandan Road 80, Shanghai 200030, China
        \and
        University of Chinese Academy of Sciences, Beijing 100049, China
        \and
        Department of Astronomy, Tsinghua University, Beijing 100084, China
        }
        
  \date{Received xxx; accepted xxx}

 
  \abstract
  {We present a Simulation-Based Inference (SBI) framework for cosmological parameter estimation via void lensing analysis. Despite the absence of an analytical model of void lensing, SBI can effectively learn posterior distributions through forward modeling of mock data. We develop a forward modeling pipeline that accounts for both cosmology and the galaxy-halo connection. By training a neural density estimator on simulated data, we infer the posteriors of two cosmological parameters, $\Omega_m$ and $S_8$. Validation tests are conducted on posteriors derived from different cosmological parameters and a fiducial sample. The results demonstrate that SBI provides unbiased estimates of mean values and accurate uncertainties. These findings highlight the potential to apply void lensing analysis to observational data even without an analytical void lensing model.}


  \maketitle
%

\section{Introduction} \label{sec:intro}

The Large Scale Structure of dark matter contains fruitful information about our Universe. Since dark matter does not interact with photons, its large scale distribution should be traced by visible proxies like galaxies. By measuring the statistics of galaxies such as the clustering of galaxies and the distortions of galaxy shapes (caused by gravitational lensing), one can analyze the underlying distributions of dark matter and make constraints on the $\Lambda$CDM model, which by now provides the best description of our Universe. Recent observational data, including spectroscopic surveys like Sloan Digital Sky Survey (SDSS, \citet{boss-1,boss-2}) and Dark Energy Survey Instrument (DESI, \citet{desi-1, desi-2}), and photometric surveys like Kilo-Degree Survey (KiDS, \citet{clu-lens}), Dark Energy Survey (DES, \citet{des}) and Hyper Suprime-Cam (HSC, \citet{hsc}) have made the constraints on the cosmological parameters to unprecedented precision. Recently, two new results from the latest released data have provided fresh insights into our understanding of the Universe., one is from DESI DR2 BAO measurements (with CMB and supernovae measurements) \citep{desi-dr2-bao}, which shows a $\sim3\sigma$ deviation from $\Lambda$CDM model, preferring an evolving dark energy rather than a cosmological constant; and the other from the cosmic shear analysis with the latest KiDS-Legacy data \citep{kids-legacy-shear}, showing that the cosmological parameter $S_8=\sigma_8\sqrt{\Omega_m/0.3}$ is in agreement with results from Planck CMB. The upcoming next-generation photometric surveys like Euclid \citep{euclid}, Legacy Survey of Space and Time (LSST, \citet{lsst}) and China Space Station Telescope (CSST, \citet{csst,csst2}), and spectroscopic surveys like MUST \citep{must}, are expected to further improve the constraints and enhance our knowledge of the Universe with the help of the deeper and wider survey areas and higher data quality.

Basic cosmological data analysis relies on analyzing ``summary statistics'' compressed from raw observational data, such as power spectrum $P(k)$. By applying the data compression one can greatly reduce the dimension of the data vector, and it is easier to deduce the theoretical models for the compressed summary statistics. From now on, two point (2pt) statistics are the most used summary statistics in cosmological analysis, which contain complete information of a Gaussian random field. Since large scale cosmological fields such as dark matter density field are widely approximated as Gaussian fields, most of the cosmological information has been included in the 2pt statistics. In recent decades this analysis framework has been confirmed to be of great success by large amounts of literature \citep{pk-bk,clu-lens,cmb-lens-1,cmb-lens-2,ang-2pts,EG}. 

In recent years, with the great development of observational data, it has become increasingly appealing to extract non-Gaussian information from cosmological fields. A variety of analysis techniques beyond 2pt statistics have been developed in recent decades and have been proven to provide better constraints on the cosmological model \citep{3pcf, mcf, bk, density-split, peak, persist-homology, wst}. Void statistics, such as void abundance (the number density as a function of void size $R_v$), void clustering, and void lensing are one class of statistics that is beyond traditional 2pt statistics. Voids are large scale underdense regions in the cosmic web, filling most of the space of the universe. Due to its underdensity characteristics, dark matter structures in voids are less affected by the non-linear growth. Meanwhile, 2pt statistics of voids can contain information encoded in galaxy high order statistics, since when we correlate two voids we actually correlating $2N$ galaxies if we use $N$ galaxies to define a void. These advantages encourage recent researchers to use voids to study the distributions and evolutions of dark matter \citep{void-wp, void-info, void-cosmo}, the mass of neutrino \citep{void-nv-1, void-nv-2} and the nature of gravity \citep{void-gravity-0, void-gravity-1, void-gravity-2}. 

However, to shed light on the large scale structure of dark matter by use of voids, one meets some challenges. First, it is a challenge to build an accurate model for some void statistics. Particularly, there are few models to accurately describe the void lensing statistics, the cross-correlation between void position and shapes of background galaxies\footnote{In \citet{euclid-vl} they build a model with linear void bias to describe the void-matter cross power spectrum, but it is known that void bias is scale-dependent and has an oscillation feature on the high $k$ end due to void exclusion effects \citep{dive, v-ex}.}. The diversity of definitions of void also brings many difficulties in void statistics modeling. Another problem of void statistics is that void is identified based on the galaxy catalog, therefore in addition to cosmology, the galaxy formation process should also influence void statistics. The cosmological structure formation and the galaxy-halo connections should be simultaneously considered in the inference process in order to obtain an unbiased estimation of cosmological parameters. Both of those challenges call for new techniques. 

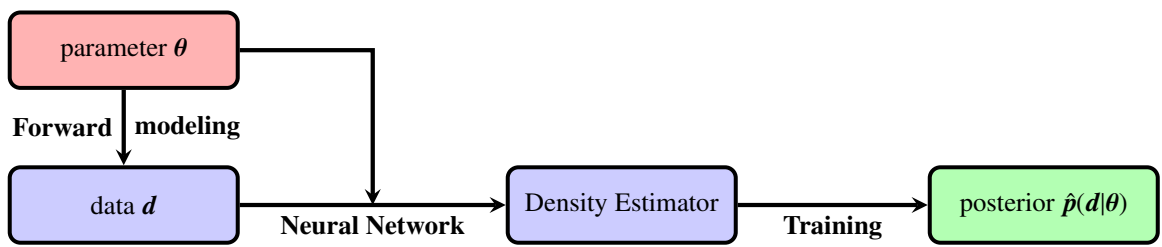
\begin{figure*}
    \centering
    \begin{tikzpicture}[node distance=2cm]
        \tikzstyle{start} = [rectangle, rounded corners, minimum width=3cm, minimum height=1cm, text centered, line width=1.5pt, draw=black, fill=red!30]
        \tikzstyle{process} = [rectangle, rounded corners, minimum width=3cm, minimum height=1cm, text centered, line width=1.5pt, draw=black, fill=blue!20]
        \tikzstyle{result} = [rectangle, rounded corners, minimum width=3cm, minimum height=1cm, text centered, line width=1.5pt, draw=black, fill=green!30]
        \tikzstyle{arrow} = [ultra thick,->,>=stealth]

        \node (param) [start] {parameter $\bm \theta$};
        \node (data) [process, below =1cm of param] {data $\bm d$};
        \node (nde) [process, right =3.5cm of data] {Density Estimator};
        \node (post) [result, right =2.5cm of nde] {posterior $\hat{\bm p}(\bm d|\bm \theta)$};

        \draw [arrow] (param) -- (data) 
                node [midway, left] {\textbf{Forward}}
                node [midway, right] {\textbf{modeling}}
                ;
        \draw [arrow] (param) -| ($(data.east)!0.5!(nde.west)$);
        \draw [arrow] (data) -- (nde)
                node [midway, below] {\textbf{Neural Network}}
                ;
        \draw [arrow] (nde) -- (post)
                node [midway, below] {\textbf{Training}}
                ;
    \end{tikzpicture}
    \caption{The workflow of a typical SBI, in which the density estimator is chosen as the Neural Posterior Estimator (NPE), i.e., directly estimating the posterior. The forward modeling process transforms parameters to be inferred to data vector, and therefore includes the process of measuring summary statistics from simulation results in our case. The density estimator takes parameter-data pair $\{\bm \theta, \bm d\}$ as input, and after the training procedure returns an estimation of posterior.}
    \label{sbi_routine}
\end{figure*}

Recently, Machine Learning (ML) techniques become popular in astrophysics and cosmology. With the help of ML, Simulation-Based Inference (SBI) \footnote{which is also called Implicit Likelihood Inference (ILI) and Likelihood-Free Inference (LFI)} has been developed to deal with cosmological inferences. The principle of SBI is that, instead of constructing an explicit form of likelihood, one can train a neural network to learn the likelihood directly from the simulation data. Therefore, SBI can be applied to any statistics that can be \textit{simulated}, regardless of whether it has a theoretical model. Plenty of successful applications of SBI in different fields of cosmology have been presented. For example in weak lensing studies, \citet{kids-sbi} constructs a systematical analysis pipeline for the analysis of cosmic shear for KiDS-1000 survey; \citet{des-sbi} presents SBI analysis to infer $w$CDM model parameters using DES-Y3 data; \citet{hsc-sbi} uses SBI to analyze HSC-Y1 weak lensing data to explore the performance of high-order statistics. Another example is the SIMulation-Based Inference of Galaxies (\textsc{SimBIG}) project \citep{simbig-fligc, simbig-fwdm, simbig-mock, simbig-mpk, simbig-nG, simbig-nlb, simbig-skew, simbig-wst}, which presents series works to construct the SBI analysis pipeline to make cosmological constraints on cosmology using galaxy clustering. In particular, they simultaneously consider the cosmological and galaxy-halo connection (Halo Occupation Distribution, HOD) models, which provide a significant reference in studying galaxy-based cosmological probes. SBI is also applied in other fields of cosmology including research on strong lensing, the epoch of reionization, and neutrino mass \citep{sbi-SL, sbi-eor, sbi-mnu}.

In this work, we apply SBI on the void lensing statistics and explore the constrain power on cosmological parameters. Motivated by the series work of \textsc{SimBIG}, we build a similar workflow of the inference, but with almost all the different details. First, due to the large scale sensitivity of void lensing signals, we use fast Particle-Mesh simulation rather than high-fidelity N-body simulation to construct the training set and validation set, which can make a larger amount of training sets with a relative small cost. Besides, the galaxy-halo connection is taken into account through the Halo Abundance Matching (HAM) technique \citep{sham0, sham1, sham2}. Compared with HOD, HAM has fewer model parameters which simplify the training process. Finally, our statistics are different from the galaxy clustering. The result will be complementary to the presented constraining results.

The rest of the paper is organized as follows. Sec.~\ref{sec:method} introduces the Simulation-Based Inference, including the motivation and workflow. In Sec.~\ref{sec:forward_model} we describe our forward modeling pipeline in detail, in which we also introduce void lensing statistics, the main statistics considered in this work. After presenting the training and validation procedures in Sec.~\ref{sec:train_val}, we finally establish our results in Sec.~\ref{sec:result}. Conclusions and future plans are shown in Sec.~\ref{sec:sum&conclu}.

\section{Method: Simulation-Based Inference} \label{sec:method}

\subsection{Motivation} \label{motivation}
In Bayesian inference, we are concerned with the posterior probability distribution $p(\bm \theta | \bm d)$ of model parameters $\bm \theta$ given some observed data $\bm d$. From the Bayes Theorem, the posterior can be written as:
\begin{equation}
    p(\bm \theta|\bm d)=\frac{p(\bm d|\bm \theta)p(\bm \theta)}{p(\bm d)}, \label{bayes_theorem}
\end{equation}
where $p(\bm d|\bm \theta)$ is called likelihood function, $p(\bm \theta)$ is the prior distribution of parameter $\bm \theta$ and $p(\bm d)$ is the Bayesian evidence which is useful in model selection tasks. Through eq.~\ref{bayes_theorem}, the estimation of posterior $p(\bm \theta|\bm d)$ can be transformed to the estimation of likelihood $p(\bm d|\bm \theta)$. In practice, the likelihood is usually chosen as a Gaussian distribution, with the mean of the data vector and the covariance of the data covariance $\bm C$. Assume the model is a mapping $f:\bm\theta \rightarrow \bm d$, the likelihood can be expressed as:
\begin{equation}
    p(\bm d|\bm\theta) \propto \exp{\left[\frac{1}{2}(\bm d-f(\bm\theta))^T \bm C^{-1} (\bm d-f(\bm\theta))\right]}.
\end{equation}
The best-fit values of parameters can be obtained by the mean/median of the posterior, which can be estimated by Monte Carlo sampling like Markov Chain Monte Carlo (MCMC).

The procedure above may fail when the calculation of the model $f(\bm\theta)$ is not efficient or even inaccessible. In other words, our aim is to perform parameter inference in the context of an implicit likelihood. This motivates the Simulation-Based Inference (SBI) technique. Instead of an explicit formula of the likelihood, SBI only requires a likelihood that we can \textit{simulate}. ``Simulate'' means a forward process that transforms the parameters $\bm \theta$ to data vector $\bm d$, and in cosmology, this transformation can usually be achieved by running cosmological simulations. Mathematically, this process actually samples parameter-data pairs $\{\bm \theta,\bm d\}$ from joint probability distribution $p(\bm \theta,\bm d)\propto p(\bm d|\bm\theta)p(\bm\theta)\propto p(\bm\theta|\bm d)$. SBI can then learn the posterior $p(\bm \theta|\bm d)$ or likelihood $p(\bm d|\bm \theta)$ from these samples with neural networks. In practice, SBI constructs the so-called density estimators\footnote{``Density" means it estimates a probability density function rather than some deterministic values.} $q_{\phi}(\bm\theta|\bm d)$, and they will be optimized to approximate true underlying posterior or likelihood (which we refer to as \textit{target distribution}) based on simulations. The $\phi$ appeared in the estimator represents free parameters to be optimized\footnote{According to the maximum-likelihood estimation principle, these parameters should be chosen when the value of $q_{\phi}(\bm\theta|\bm d)$ is maximum in each of the observation $\bm d_i$, which is the theoretical support of the training strategy}. Then, it is possible to draw samples from this estimator, as can be done in traditional Bayesian inference, and analyze the mean and covariance of the model parameters as well.

\subsection{Workflow}\label{workflow}
The basic workflow of SBI can be shown in Fig.~\ref{sbi_routine}: the first step is forward modeling, which transforms parameters to data vectors; then by use of parameter $\bm\theta$ and data $\bm d$ we can construct the neural density estimator $q_{\phi}(\bm \theta|\bm d)$. After training the neural estimator, one can obtain an estimated likelihood/posterior. Accurate modeling is essential to obtain an accurate posterior estimator, and appropriate network structures and training strategies are helpful to optimize the estimator. 

The key point of SBI is the construction of density estimators as they will be proxies of underlying true posteriors. Based on the type of target distributions, density estimators can be divided into Neural Posterior Estimator (NPE, directly estimate the posterior $p(\bm\theta|\bm d)$), Neural Likelihood Estimator (NLE, estimate the likelihood $p(\bm d|\bm\theta)$) and Neural Ratio Estimator (NRE, estimate the ratio of posterior and prior $p(\bm\theta|\bm d)/p(\bm\theta)$) in general. In this work, we use NPE to achieve our inference. Since the posterior is accessible, it is not necessary to run MCMC sampling to obtain the final constrain contour of parameters, which is highly computationally efficient.

There are several existing networks to construct density estimators, two of which are Mixture Density Network (MDN) and Masked Autoregressive Flows (MAFs). MDN can be intuitively understood as a fitting procedure that fits the target posterior by many Gaussian components with different means and covariances, and one can train a neural network to learn a set of appropriate means and covariances. The principle of MAFs is based on the chain rule of probability demonstrating that any joint probability distribution $p(\bm x)$ can be decomposed into productions of a series of 1D conditional probability distributions as $p(\bm x)=\prod_i p(x_i|\bm x_{1:i-1})$. If the beginning of the chain is a standard normal distribution and each of the conditional probabilities is invertible, then this can be regarded as a Normalization Flow from $p(\bm x)$ to a standard normal distribution, which makes $p(\bm x)$ easy to sample. As a density estimator, MAFs actually learn these conditional probabilities by some parameterized functions $q_{\phi}(x)$. In practice, one can either use one kind of network architecture or use an ensemble of different architectures to construct the density estimators so as to achieve better performance.

After determining the density estimator, the next step is to prepare training and validation sets. This is another key process in SBI since all of the information is learned from the training sets. In the following part we introduce our forward modeling pipeline, in which we start with the dark matter field and finally obtain our summary statistics.

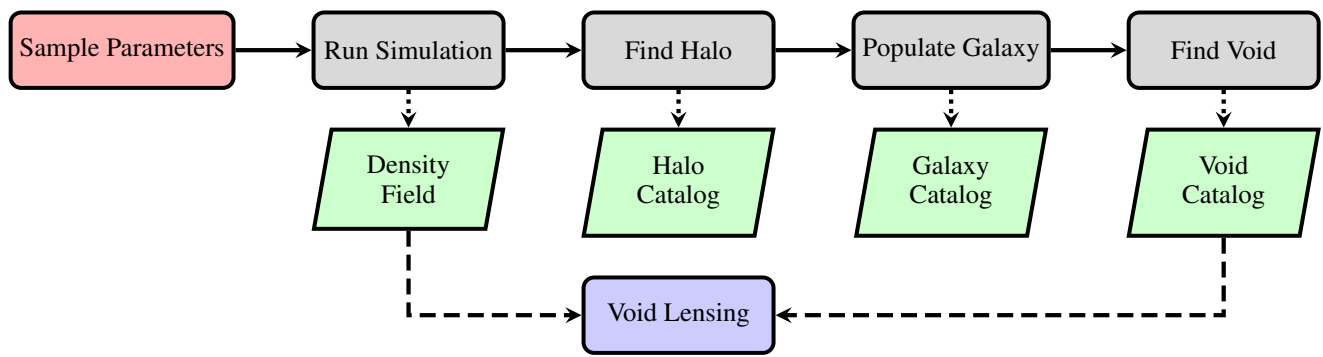
\begin{figure*}
    \centering
    \begin{tikzpicture}[node distance=2cm]
        \tikzstyle{start} = [rectangle, rounded corners, minimum width=2.5cm, minimum height=1cm, text centered, line width=1.5pt, draw=black, fill=red!30]
        \tikzstyle{process} = [rectangle, rounded corners, minimum width=2.5cm, minimum height=1cm, text centered, line width=1.5pt, draw=black, fill=gray!30]
        \tikzstyle{output} = [trapezium, trapezium left angle=80, trapezium right angle=100, minimum width=2.5cm, minimum height=1cm, text centered, line width=1.5pt, draw=black, fill=green!20, align=center]
        \tikzstyle{result} = [rectangle, rounded corners, minimum width=2.5cm, minimum height=1cm, text centered, line width=1.5pt, draw=black, fill=blue!20]
        \tikzstyle{arrow} = [ultra thick,->,>=stealth]
        \tikzstyle{darrow} = [ultra thick,->,dotted,>=stealth]

        \node (param) [start] {Sample Parameters};
        \node (simulation) [process, right =1cm of param] {Run Simulation};
        \node (DM) [output, below =0.5cm of simulation] {Density\\ Field};
        \node (findhalo) [process, right =1cm of simulation] {Find Halo};
        \node (halocat) [output, below =0.5cm of findhalo] {Halo\\ Catalog};
        \node (findgal) [process, right =1cm of findhalo] {Populate Galaxy};
        \node (galcat) [output, below =0.5cm of findgal] {Galaxy\\ Catalog};
        \node (findvoid) [process, right =1cm of findgal] {Find Void};
        \node (voidcat) [output, below =0.5cm of findvoid] {Void\\ Catalog};
        \node (vlens) [result, below =0.5cm of halocat] {Void Lensing};

        \draw [arrow] (param) -- (simulation);
        \draw [arrow] (simulation) -- (findhalo);
        \draw [darrow] (simulation) -- (DM);
        \draw [arrow] (findhalo) -- (findgal);
        \draw [darrow] (findhalo) -- (halocat);
        \draw [arrow] (findgal) -- (findvoid);
        \draw [darrow] (findgal) -- (galcat);
        \draw [darrow] (findvoid) -- (voidcat);
        \draw [ultra thick, ->, dashed, dash pattern=on 6pt off 3pt, >=stealth] (voidcat) |- (vlens);
        \draw [ultra thick, ->, dashed, dash pattern=on 6pt off 3pt, >=stealth] (DM) |- (vlens);
    \end{tikzpicture}
    \caption{The schematic diagram of forward modeling. The gray blocks represent modeling processes, the green blocks represent outputs in each step of modeling. The red block represents input parameters in the model. The gray blocks represent steps in forward modeling, and the green blocks show the corresponding output in each step. Final output i.e. void lensing signal is shown in blue block.}
    \label{fwd_model}
\end{figure*}

\section{Forward Modeling} \label{sec:forward_model}
In this section, we describe our forward modeling pipeline in detail. Fig.~\ref{fwd_model} shows the flowchart of our pipeline. At the beginning we input the cosmological parameters, generate the initial dark matter distributions, and run simulations to obtain the dark matter field at the present time. We then apply the halo finder algorithm on the dark matter field to identify dark matter halos.
Galaxies are then populated to halos through Halo Abundance Matching (HAM) method. Voids are identified in the galaxy catalog, and finally we measure the void lensing signals.

\begin{figure}
    \centering
    \includegraphics[width=0.8\linewidth]{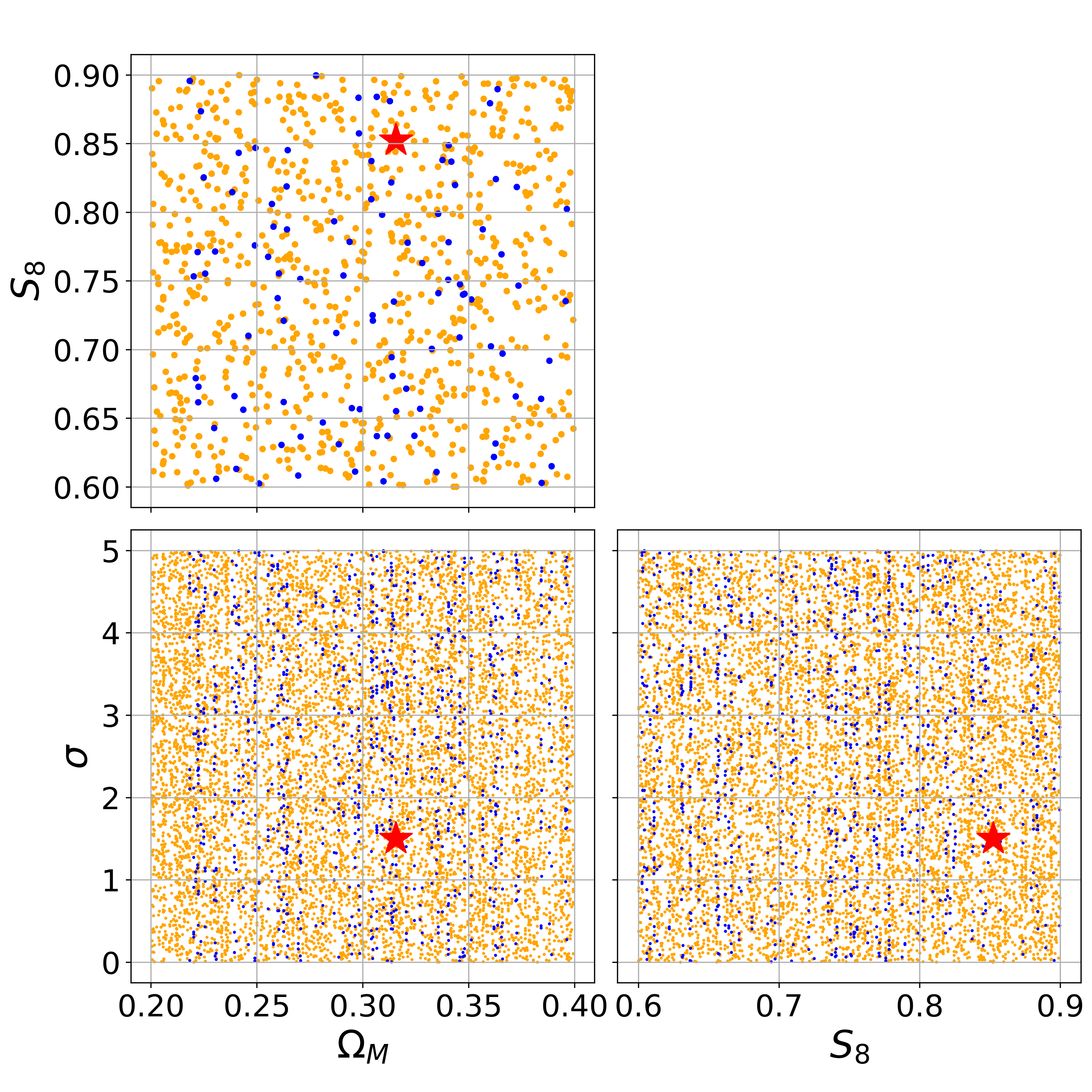}
    \caption{Parameter space of parameters considered in this work, including cosmological parameters $\Omega_M$, $S_8$, and HAM parameter $\sigma$. These samples are randomly divided into two parts with $9:1$, corresponding to the training set (orange dots) and validation set (blue dots), respectively. In order to test the performance of our method on a special cosmology, we choose a fiducial cosmology marked as red star. Note this sample is neither in the training set nor validation set.}
    \label{param_space}
\end{figure}

\subsection{FastPM simulation} \label{fpm}
\textsc{fastpm}\footnote{\url{https://github.com/fastpm/fastpm}} \citep{FastPM} is a Particle-mesh (PM) simulation that trade-off the accuracy for speed as compared to N-body simulations. It maps the particles to mesh grids to estimate the gravitational force, therefore loss accuracy on scales smaller than grid size. Besides, it modifies the traditional kick and drift factor to ensure the correctness of linear displacement evolution on large scales regardless of the number of time-steps. 

In this work, we have run \textsc{fastpm} with $1024^3$ dark matter particles in a $1~h^{-1}\text{Gpc}$ box. The force resolution $B$, which is the ratio between mesh size and number of particles, is chosen as $B=2$, and 40 linear steps are used to evolve the density field from $a=0.01$ to $a=1$.

Because of the limitation of computing resources, in this work we only consider two cosmological parameters that are expected to be sensitive to lensing statistics: the matter density parameter, $\Omega_M$ and the standard deviation of matter density fluctuation in $8~h^{-1}\text{Mpc}$, $\sigma_8$. Due to the well-known degeneracy between these two parameters, for example the ``banana-shaped'' contour on $\Omega_M-\sigma_8$ plane for cosmic shear statistics, we choose $S_8=\sigma_8\sqrt{\Omega_M/0.3}$ to replace the original $\sigma_8$ parameter. Thus, the cosmological parameter space is constructed by $\Omega_M$ and $S_8$. We sampling independent 1000 parameter pairs from the priors $\Omega_M\sim\mathcal{U}(0.2, 0.4)$ and $S_8\sim\mathcal{U}(0.6, 0.9)$, where $\mathcal{U}$ represents uniform distribution. The upper left of fig.~\ref{param_space} provides a visualization of distributions of cosmological parameter samples. These 1000 parameters are then as the input of \textsc{fastpm} to generate dark matter fields. In order to consider the variety in $p(\bm\theta, \bm d)$ along dimensions of data $\bm d$, we use different random seeds for each cosmological simulation. 

\subsection{Galaxy and Void Catalogs} \label{gal_void}

We first construct the dark matter halo catalog. We choose \textsc{rockstar} \citep{rockstar} software to find halos. This algorithm considers 6-D phase space information to identify halos, which is tested in some previous works (e.g., \citet{void-nv-2}) that performs well in \textsc{fastpm}.

Voids can be identified in either halo catalogs or galaxy catalogs. Since halos cannot be observed directly in real observation, it is better to choose galaxies as tracers of voids. This means galaxy-halo connection should be considered in the modeling. Halo Abundance Matching (HAM) is a simple and intuitive empirical method to describe the galaxy-halo connection. It is based on the assumption that there is a monotonic relation (not necessarily linear) between the galaxy luminosity (or stellar mass) and the halo mass (or the circular velocity). Besides, to model the stochasticity in the galaxy–halo mass relation \citep{hm-relation1,hm-relation2} we consider a Gaussian scatter with dispersion $\sigma$ (in the following we will call it ``HAM parameter'') to the halo mass \citep{sham-disp}. In practice, we first multiply $S_g$ to the halo mass $M_h$ to obtain $M_{\text{scat}}=M_h\times S_g$ for each halo in the simulation, where $S_g=1+\mathcal{N}(0,\sigma^2)$ when $\mathcal{N}(0,\sigma^2)>0$ and $S_g=\exp\left(\mathcal{N}(0,\sigma^2)\right)$ otherwise, and sort these halos in descending order of $M_{\text{scat}}$, then populate a galaxy in the center of each halo in the catalog from the most massive ones to the least ones until we get the expected number of HAM galaxies $N_{\text{gal}}=n_{\text{ref}}V_{\text{box}}$.

In principle, for each halo catalog one can fit some statistics (e.g., projected 2 point correlation function $w_p$) to the real data in order to obtain an optimized $\sigma$ corresponding to a realistic galaxy mock. However, simultaneously optimizing cosmological and HAM parameters is available for SBI, which is a more self-consistent method. Meanwhile, purely using SBI to analyze cosmology and galaxy-halo connection can naturally consider the correlations between cosmological and HAM parameters. Therefore, we set HAM parameter $\sigma$ to be a free parameter and try to train SBI to extract the influence from galaxy formation processes on void lensing signals. Following \citet{simbig-mock}, we randomly sample 10 HAM parameters from prior $\mathcal{U}(0, 5)$ for each cosmology, and finally we have 10,000 mock galaxy catalogs in total. The reference number density of galaxies is chosen as $n_{\text{ref}}=3.5\times 10^{-4}~(h/\text{Mpc})^3$ to match the BOSS LOWZ LRG samples \citep{boss-2}.

Now it is ready to identify voids on the galaxy catalog. We use \textsc{dive}\footnote{\url{https://github.com/cheng-zhao/DIVE}} \citep{dive} to construct our void catalog. \textsc{dive} applies Delaunay Triangulation on the galaxy catalog, and defines voids as the empty circumspheres constrained by tetrahedra of galaxies. The void center and radius are regarded as those of sphere. These voids are prepared for measurements of void lensing signals in the next step.

\begin{figure*}
    \centering
    \includegraphics[width=1\linewidth]{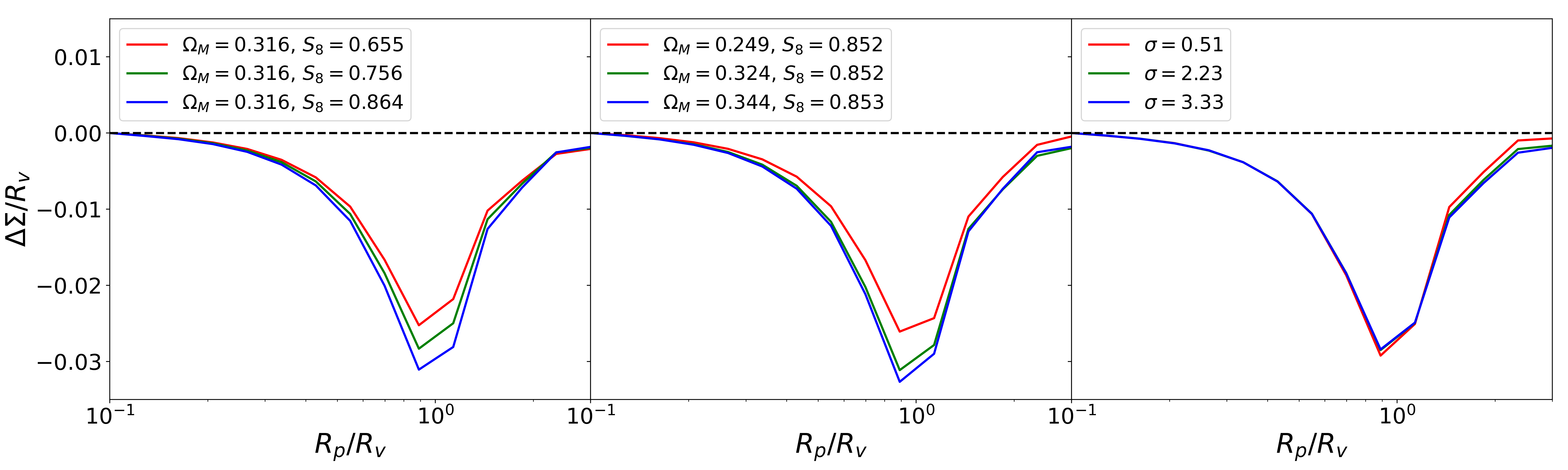}
    \caption{The void lensing signals in different cosmologies and different galaxy population processes. The x-axis represents the projection radius rescaled by the void radius. The y-axis is the Excess Surface Density (ESD) around voids. These are the data vectors used by SBI.}
    \label{signal}
\end{figure*}

\subsection{Void Lensing Measurement} \label{measure}
The gravitational field of the foreground matter will bend the light rays emitted from the distant galaxies, which will change their luminosity and shape. The distortion of the galaxy shape can be quantified by tangential shear $\gamma_t$. It is related to the so-called lensing convergence $\kappa$:
\begin{equation}
    \gamma_{\text{t}}=\bar{\kappa}-\kappa, \label{gt_kappa}
\end{equation}
where $\kappa$ is a weighted integral over comoving distance of matter density $\delta_m$, which can be written as:
\begin{equation}
    \kappa = \frac{3\Omega_m H_0^2}{2}\int{\dd \chi \frac{\chi_l(\chi_s - \chi_l)}{\chi_s}\frac{\delta_m}{a_l}}, \label{kappa_dm}
\end{equation}
and $\bar{\kappa}=(2/R^2)\int_0^R{\dd R^{\prime} R^{\prime}\kappa(R^{\prime})}$ is the mean convergence inner the sphere with radius R. If we define the critical surface density $\Sigma_{\text{crit}}$:
\begin{equation}
    \Sigma_{\text{crit}}=\frac{a_l}{4\pi G}\frac{\chi_s}{\chi_l (\chi_s-\chi_l)}, \label{sig_crit}
\end{equation}
and further assume that the foreground matter density is associated with an isolated object (which in this work is an isolated void), eq.~\ref{kappa_dm} can be:
\begin{equation}
    \kappa=\frac{1}{\Sigma_{\text{crit}}}\bar{\rho}_m\int{\dd\chi \delta_m}=\frac{\Sigma}{\Sigma_{\text{crit}}}. \label{kappa_dm_2}
\end{equation}

From eqs.~\ref{gt_kappa} and \ref{kappa_dm_2} one can see that by analyzing the distortion of the background galaxies it is possible to catch up information of foreground matter density distribution $\delta_m$. Similar to galaxy-galaxy lensing, in which case the foreground objects are galaxies, void lensing describes the light distortion effects when the foregrounds are voids. Due to the underdensity characteristic of voids, void lensing will provide a ``minus'' signal compared with galaxy-galaxy lensing.

In observation one often choose the excess surface mass density $\Delta\Sigma$ as the observable. This is related to previous quantities by:
\begin{equation}
    \Delta\Sigma=\gamma_t\Sigma_{\text{crit}}=\Sigma_{\text{crit}}(\bar{\kappa}-\kappa). \label{dsigma}
\end{equation}
In the following paper ``void lensing signal'' is another word of the excess surface mass density of voids. From eq.~\ref{kappa_dm_2} and eq.~\ref{dsigma} it can be found that the void lensing signal is just an integral transform of void profiles, and therefore in simulation it can be obtained in an inverse method, i.e., measuring the void profiles and transform to lensing signals. This can reduce a great amount of calculations compared with running a full ray-tracing simulation. In this work we choose this approximate technique and leave a realistic modeling to future work.

Measuring void profiles in simulation is equivalent to measuring the void-dark matter cross correlation. We use \textsc{pyfcfc}\footnote{\url{https://github.com/dforero0896/pyfcfc}}, which is a python-wrapper of \textsc{fcfc}\footnote{\url{https://github.com/cheng-zhao/FCFC}} \citep{fcfc} to measure the correlation function, and calculate $\Delta\Sigma$ use eqs.~\ref{kappa_dm_2} and \ref{dsigma} as void lensing signal. We first cut the voids into the void radius interval $15<R_v<25~\text{Mpc}/h$, and divide those voids into 10 linear size bins, each with a width of $\Delta R_v=1~\text{Mpc}/h$. Next, for each void size bin we measure correlation functions in 15 logarithm separation bins with a range of $r \in [0.1, 3]R_v$, where $R_v$ is the void radius. Then, these correlation functions are stacked to construct a higher signal-to-noise ratio (SNR) signal. This leads to a data vector with dimension 15. 

\begin{figure*}
    \centering
    \includegraphics[width=1\linewidth]{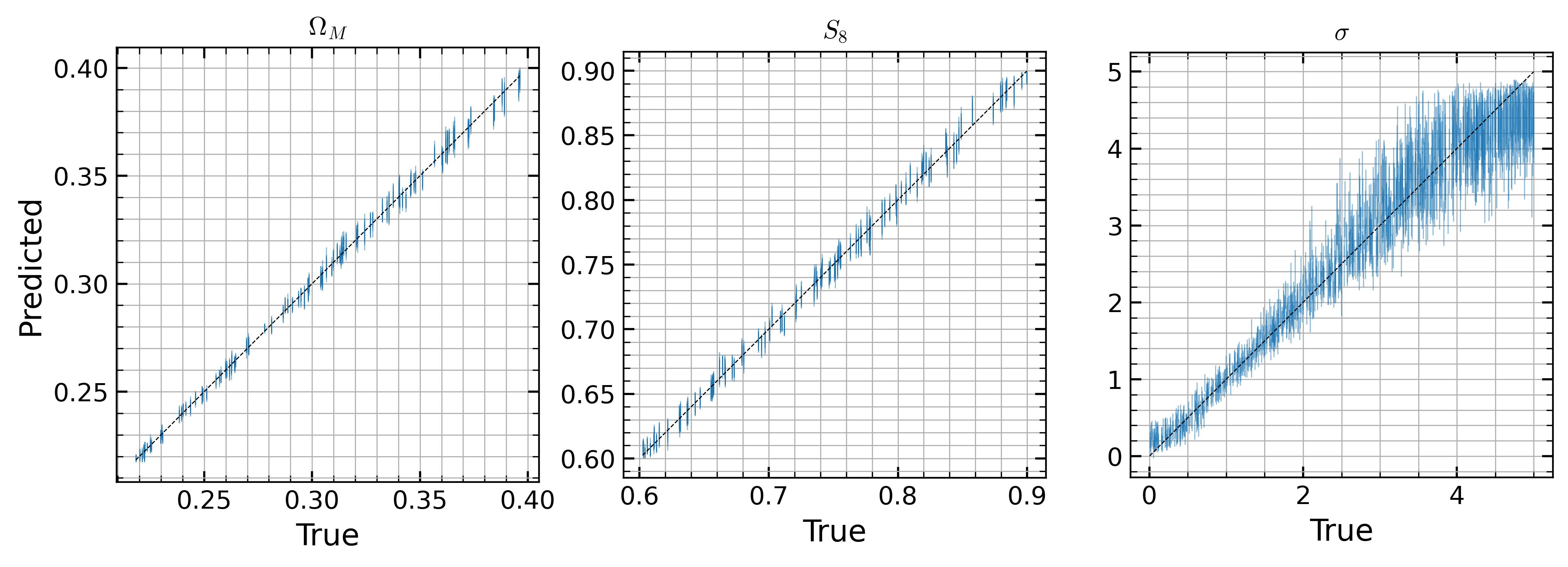}
    \caption{The predicted mean values versus the truths of cosmological parameters $\Omega_M$ and $S_8$, and HAM parameter $\sigma$, over 1,000 validation sets for each parameter. The blue sticks represent $1\sigma$ regions of predictions.}
    \label{predict}
\end{figure*}

\section{Training \& Validation} \label{sec:train_val}
After the forward modeling process, we finally have 10,000 samples for SBI (1,000 cosmologies $\times$ 10 HAMs per cosmology). These samples should first be separated into training and validation sets. $90\%$ of the total samples are regarded as training samples while the remaining are prepared for validation. Note that only 1,000 samples are independent in terms of cosmological parameters. Therefore, the ratio of the number of independent cosmology samples between the training set and validation set should also be $9:1$. Therefore we require that different HAM samples with the same cosmological parameter should not be divided into training and validation sets, but should be put into either training or validation set as a whole. In fig.~\ref{param_space} we depict the distributions of training and validation sets in orange and blue dots, respectively.

In this work we use the public code \textsc{ltu-ili}\footnote{\url{https://github.com/maho3/ltu-ili}} \citep{ltu-ili} to train our density estimator $q_{\phi}(\bm\theta|\bm d)$. The Masked Autoregressive Flows (MAFs) and Mixture Density Network (MDN) are both used to construct the density estimator. For MAFs, we use 5 MADE blocks, each of which has 50 hidden features. For MDN we use 6 components and each of them has 50 hidden features. The learning rate is set to be $1\times10^{-3}$ and the batch size is 32. In order to alleviate the stochasticity due to the random initialization of the networks, we independently train 10 rounds for each of the estimators with the same hyper-parameters, choose the best 6 results (3 MAFs, 3 MDNs), and ensemble them to construct our final estimator.

It is a necessary and critical challenging task in SBI to validate the training results. If the analytical form of the data vector is accessible, it is possible to construct an \textit{accurate} Gaussian likelihood, and one can compare the results of SBI with it. However, the Gaussianity of the likelihood also needs to be confirmed. Cases will be even worse when there is no analytical form, revealing that theoretical likelihood is inaccessible. In recent years several validation methods have been published, which is helpful for testing the accuracy of the estimators. In this work, we choose Tests of Accuracy with Random Points (TARP) as a metric to assess if our posterior estimator can neither underestimate nor overestimate the uncertainties of parameters. In brief, TARP compares the Expected Coverage Probabilities (ECPs) of random posterior samples with a given credibility level of the estimated posterior, and these two should be consistent for an accurate probability density function. The credibility level $1-\alpha$ is defined as the integration area of \textit{estimated} posterior under specific parameter intervals: $\int_{V}{\dd \theta\hat{p}(\theta|d)}=1-\alpha$, the Coverage Probabilities (CPs) is the integration area of \textit{true} posterior under the same intervals $\int_{V}{\dd \theta p(\theta|d)}=CP$, and the Expected Coverage Probabilities (ECPs) is the expectation of CPs among different data $ECP= \mathbb{E}_{p(d)}[CP]$. Theoretical and technical details can be found in \citet{tarp}.

\section{Results} \label{sec:result}
In this part, we present our main results. First, we show our data vectors, i.e. void lensing signals, and discuss their dependence on cosmological parameters and HAM parameters. Then we evaluate the performances of our estimated posterior on different cosmology samples, focusing on the validation of predictions of mean values and uncertainties. Performances on fiducial cosmology samples is evaluated in the final part.

\subsection{Void Lensing Signal}\label{data_vec}

Fig.~\ref{signal} shows the void lensing signals measured in our simulations. The x-axis represents projection radius $R_p$ rescaled by void radius $R_v$, and y-axis is the excess surface density $\Delta\Sigma$ in units of $h^2 M_{\odot}/\text{pc}^2/\text{Mpc}$. Both the two parameters $\Omega_M$ and $S_8$ can affect the amplitude of void lensing signals, and higher $\Omega_M$($S_8$) will both induce a deeper signal. This of course reveals that void lensing can be used to constrain cosmology, but infers a degeneracy of these two parameters with respect to void lensing. On the other hand, the HAM parameter $\sigma_M$ does not greatly change the amplitude of the lensing signal, but rather influences the shape of the signal in the range of $r>R_v$. This will be seen in the following constraining contour, in which the degeneracy direction of the HAM parameter and the other two cosmological parameters are almost orthogonal. The broken degeneracy of these two classes of parameters is exactly good news: considering the galaxy populations in halos may not greatly influence the constraining power of cosmology.

\subsection{Predictions of different cosmology}\label{validation}

We first validate if the predicted mean values of parameters are unbiased in different cosmologies, and if the estimated uncertainties are neither overconfident nor underconfident. For the former, we directly compare the SBI predictions of the model parameters and the truth; and for the latter, the TARP metric is applied to evaluate the accuracy of our posterior especially the estimation of the uncertainty. As is discussed in Sec.~\ref{sec:train_val}, 1,000 samples are prepared for the validation with 100 of them having independent cosmological parameters.

Fig.~\ref{predict} shows the predicted mean values as a function of the truths for all the cosmological parameters and HAM parameters. The x-axis represents the true value and the y-axis is the predicted value. For cosmological parameters, our estimator can give unbiased predictions for all of the validation samples. But for the HAM parameter, especially for the larger values ($\sigma>2.5$), the predictions are not quite well. This may be due to the fact that void lensing signals show less response for the larger $\sigma$, but may also imply that $\sigma$ is not well trained resulting from the low sampling density for $\sigma$ parameter (10 samples in $\left[0,5\right]$ for one cosmology). Since in real observation, low redshift galaxy samples prefer a small value of $\sigma$ ($\sigma<1$), it is not a severe problem in real-world applications. Besides, in this work we are only interested in the cosmological parameters and treat the galaxy-halo connection parameter as a nuisance parameter to be marginalized over, and due to the weak degeneracy between HAM and cosmological parameters (see fig.~\ref{contour}), constraints on the cosmology should not be greatly influenced.

\begin{figure}[ht]
    \centering
    \includegraphics[width=1\linewidth]{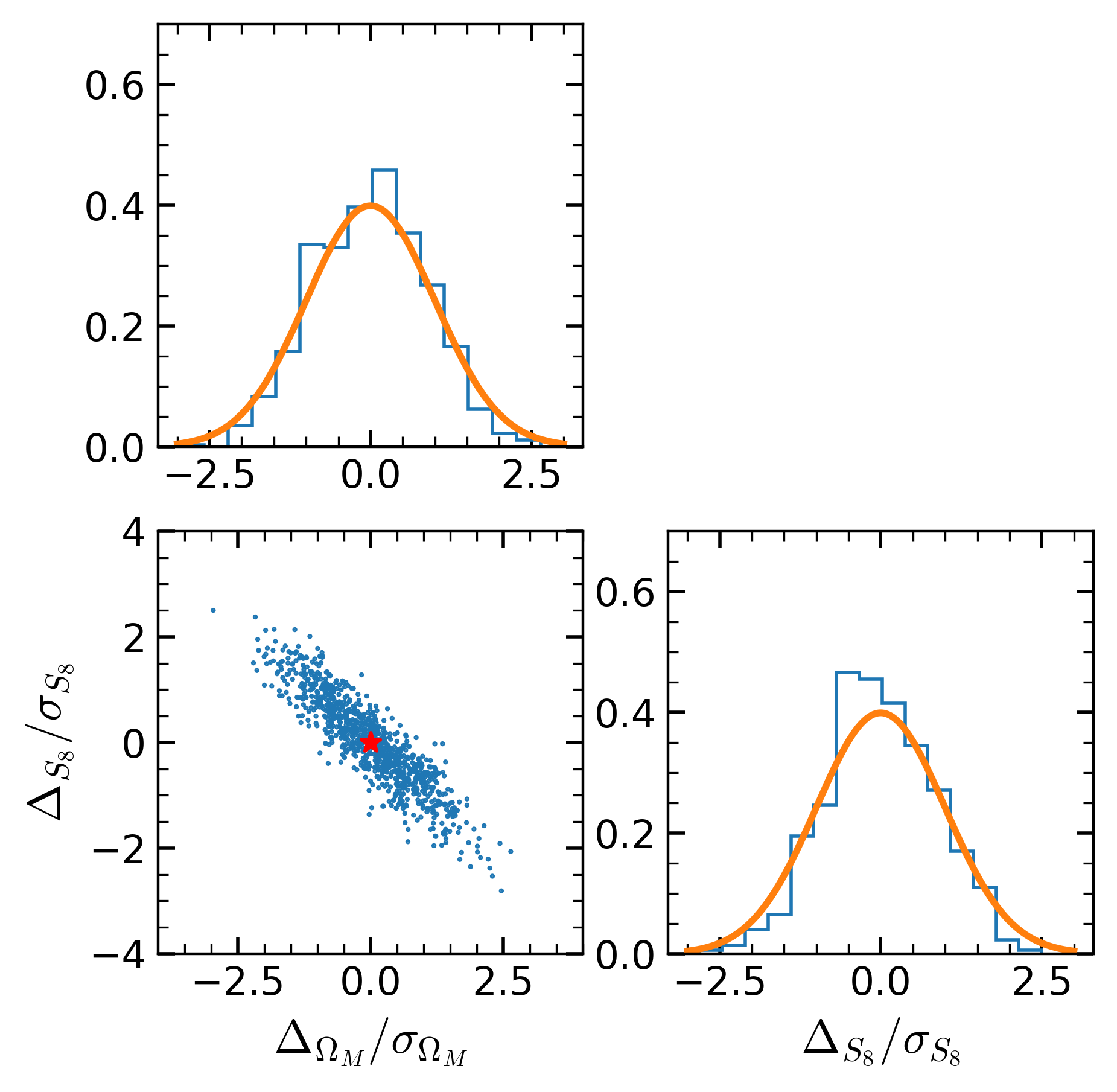}
    \caption{The differences between predicted values and truths (blue dots), shown in $\Omega_M-S_8$ plane. The red star represents the truth. We also present the corresponding 1D distribution of the two cosmological parameters, as well as a reference standard normal distribution (orange line). It can be seen that they are greatly consistent.}
    \label{predict-2d}
\end{figure}

The accuracy of the predictions of cosmological parameters is shown in fig~\ref{predict-2d}. We calculate the differences between predicted values and truths relative to their $1\sigma$ uncertainties, $\Delta_X/\sigma_X$, in which $X=\Omega_M, S_8$, and show distributions of these values. As shown in fig~\ref{predict-2d}, we find the 1D distributions (blue histograms) for both cosmological parameters to be consistent with the standard normal distribution (orange line). These two tests confirm that for cosmological parameters drawn from the prior distribution of the training set, SBI provides unbiased predictions.

\begin{figure}
    \centering
    \includegraphics[width=1\linewidth]{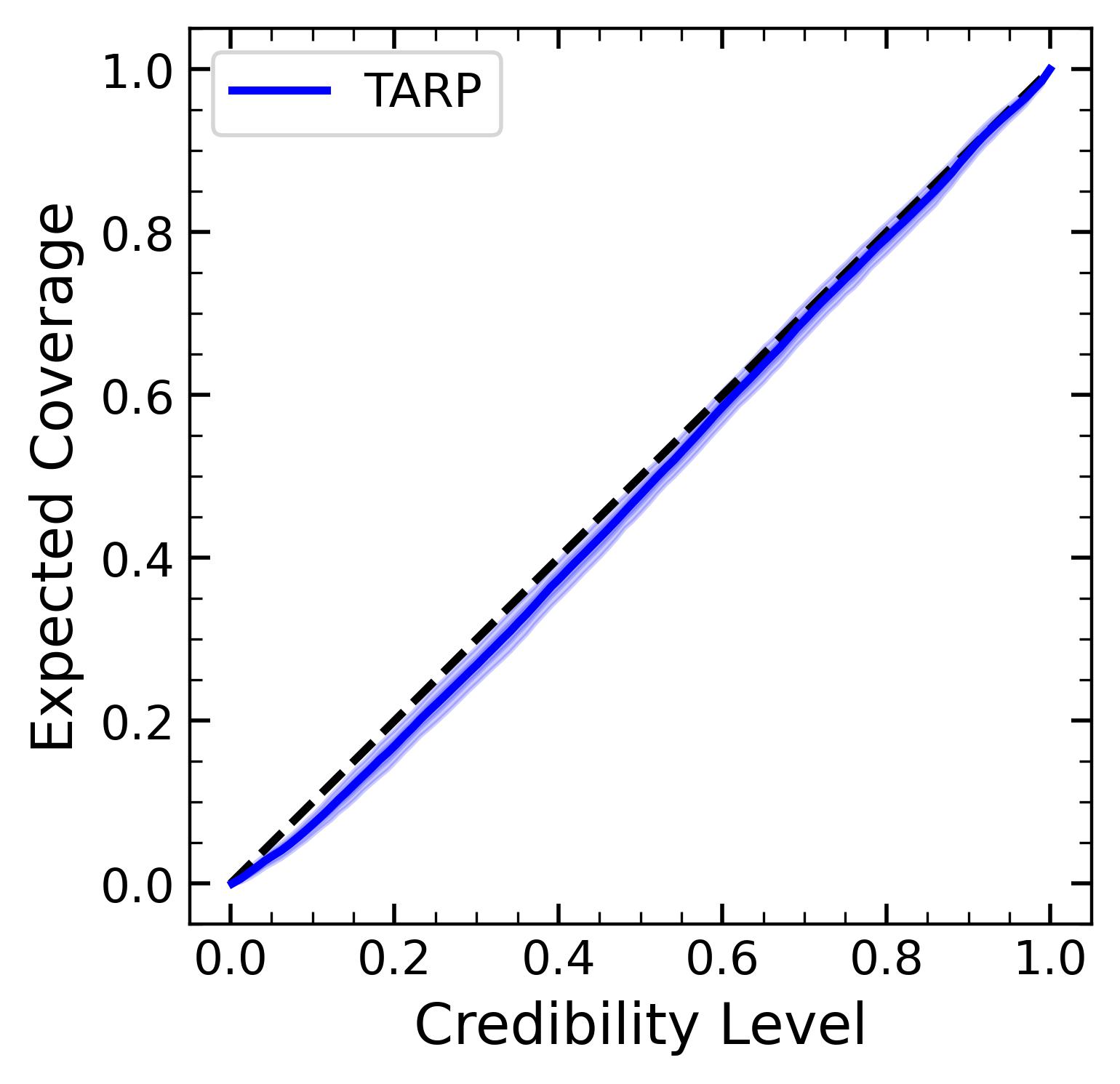}
    \caption{The Expected Coverage Probability (ECP) versus credibility level in TARP. The Expected Coverage Probability (ECP) versus credibility level in TARP. The dark blue region indicates $1-\sigma$ credible region from the bootstrapping technique from 100 realizations, while the light region indicates $2-\sigma$. If the uncertainties estimated from the posterior are accurate, the blue line should be along with the diagonal.}
    \label{tarp}
\end{figure}

Besides, the accuracy of the uncertainties estimated by SBI is also tested by the TARP metric, and the result is established in Fig.~\ref{tarp}. The blue line represents the Expected Coverage Probability (ECP) given the credibility level $1-\alpha$. If the density estimator is accurate, i.e., the estimation of the uncertainties is neither underconfident nor overconfident, ECP should be equal to $1-\alpha$, and the blue line should be along with the diagonal (black dashed line). If uncertainties are overestimated (i.e., underconfident), the TARP regions from randomly selected points are more likely to cover approximately half of the posterior estimator $\alpha\sim0.5$ \citep{tarp}, resulting in a ``S'' type curve. The light and dark shaded region of the TARP lines shows the $1\sigma$ and $2\sigma$ confidential region which is generated by 100 bootstrap samples of the validation sets. It shows that we obtain a highly accurate estimation of uncertainties. A potential bias of the uncertainty estimation is barely seen compared to the statistical fluctuations of the TARP metric.

Through these two validations, we can conclude that we obtain an unbiased amortized posterior estimator (i.e., SBI works on any cosmology with parameters drawn from the prior) and accurate uncertainties. These results support the accessibility of applying SBI on cosmological inference by void lensing.

\subsection{Validations on multiple realizations of fiducial cosmology}\label{posterior}
Now we are interested in the validation of SBI on a specific cosmology sample. We run a fiducial simulation with $\Omega_M=0.3156$ and $\sigma_8=0.831$, and set HAM parameter $\sigma=1.5$. After obtaining the void lensing signals following the same forward modeling process described in Sec.~\ref{sec:forward_model}, we apply SBI on it and obtain the predictions for these parameters.

\begin{figure}
    \centering
    \includegraphics[width=1\linewidth]{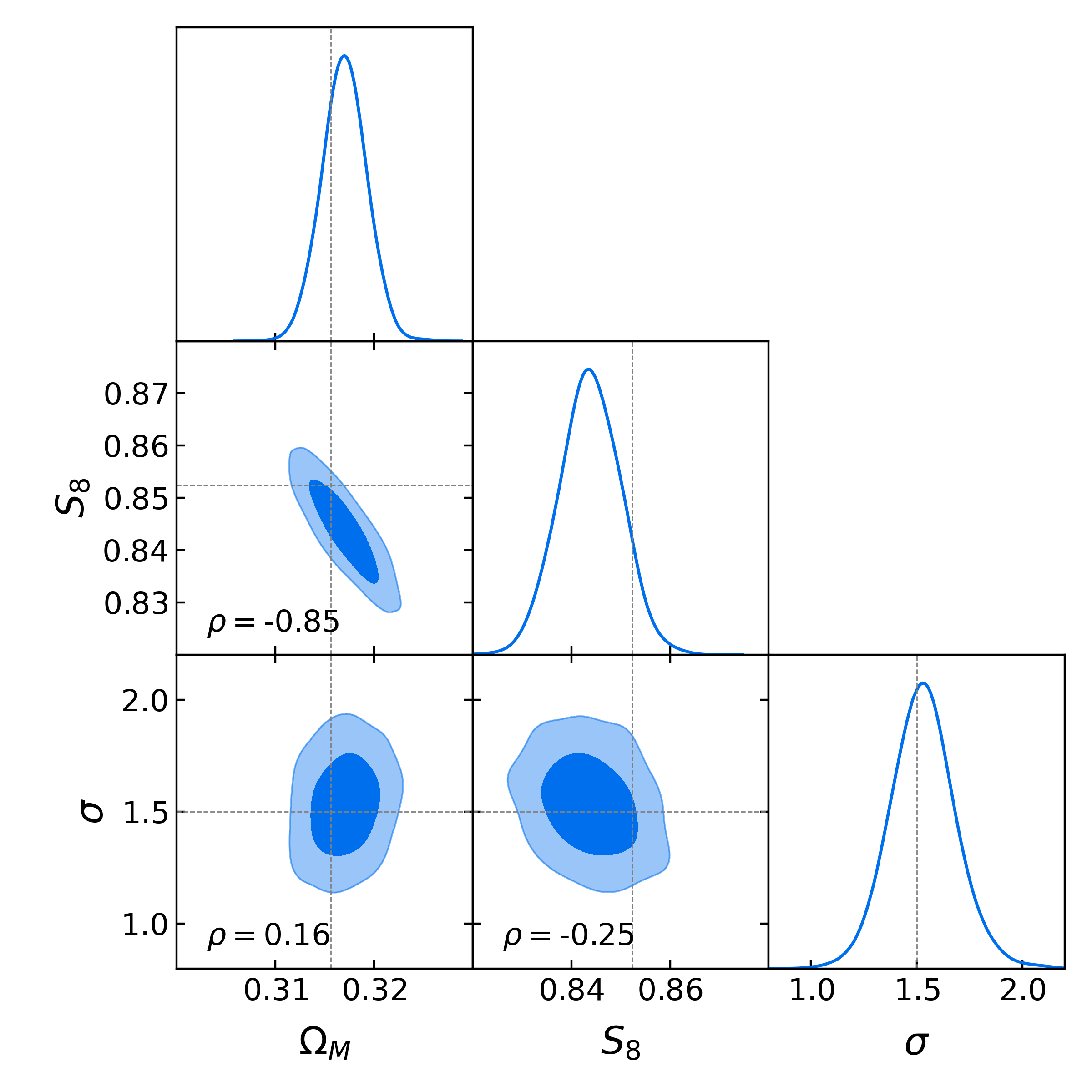}
    \caption{The posterior prediction from SBI on data vector of fiducial cosmology. The thin gray lines mark the true value and $1-\sigma$ and $2-\sigma$ regions are presented as dark and light shadows. We also mark the correlation coefficients between each pair of parameters.}
    \label{contour}
\end{figure}

Fig.~\ref{contour} establishes the constraining results as the 2D joint distributions of each pair of parameter set $\{\Omega_M, S_8, \sigma\}$ as well as 1D marginalized distributions. The gray dashed line represents the true value. The 68\% ($1\sigma$) and 95\% ($2\sigma$) credible regions are also presented in the figure with dark and light shadows. It shows that for all of the parameters, SBI can give unbiased predictions, and the largest discrepancies for the two cosmological parameters are within $1\sigma$ in this single realization.

\begin{figure}
    \centering
    \includegraphics[width=1\linewidth]{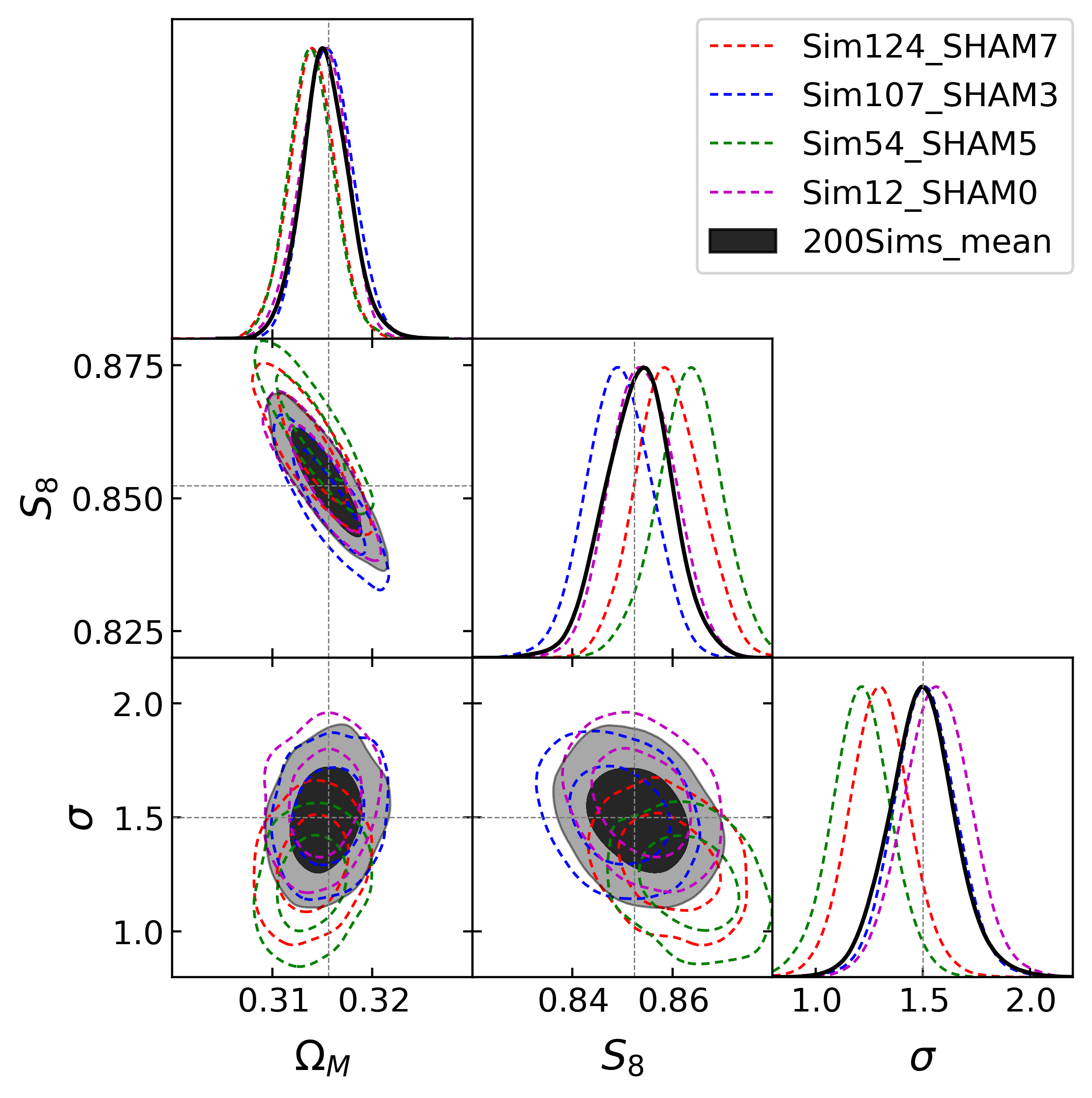}
    \caption{The same as Fig.~\ref{contour}, but for different realizations. The red, blue, green, and magenta thin contours are the predictions of four realizations randomly chosen in the validation set, and the black thick contour represents the prediction of the data vector averaged over all 200 validation samples.}
    \label{contour_mean}
\end{figure}

To figure out if this discrepancy is a result of statistical fluctuation, we further run 200 simulations with the same cosmological parameter but different initial conditions and apply SBI on these lensing signals. We then average these signals and obtain a signal with less cosmic variance. It is expected that if SBI is unbiased, it can give a more accurate prediction on this cosmic variance-reduced sample. In fig.~\ref{contour_mean} we depict the SBI prediction on the mean signal in solid filled contour, as well as predictions on four randomly chosen samples in dash empty contours. The predicted mean values for each individual sample vary around the true values, while the predictions of the mean signal are in good agreement with the true values, illustrating that SBI can provide an unbiased prediction on fiducial cosmology.
    
\begin{figure*}
    \centering
    \includegraphics[width=1.0\linewidth]{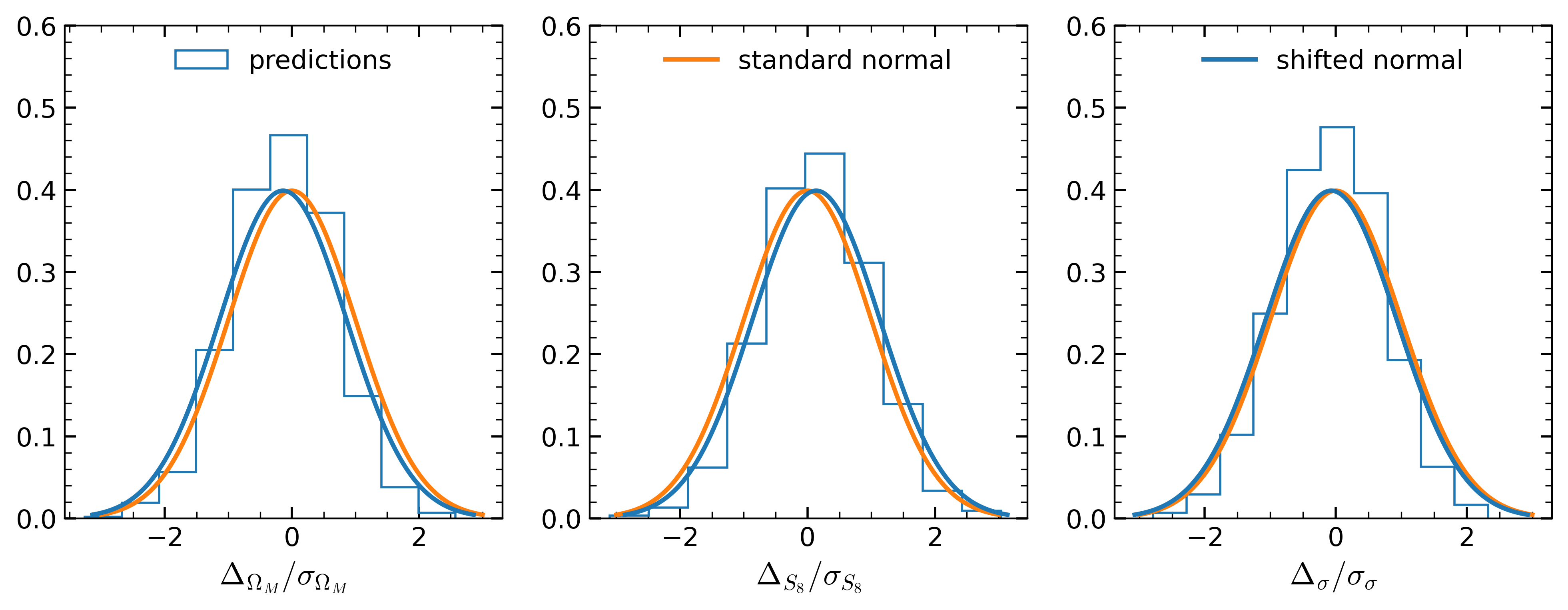}
    \caption{Distributions of SBI predictions on fiducial cosmology. For each panel, the blue histogram shows the distribution of the differences between predictions and truths, rescaled by the predicted standard deviation. The orange line represents the standard normal distribution, and the blue curve represents the normal distribution whose mean equals to the average of predictions.}
    \label{predict_rlzs}
\end{figure*}

Furthermore, we check the distribution of SBI predictions on these 200 realizations. In fig.~\ref{predict_rlzs}, we show the histogram of the differences between the predictions and the truths $\Delta_X=X_{\text{pred}}-X_{\text{true}}$ normalized by the predicted standard deviation $\sigma_X$ (where $X=\Omega_M, S_8$ or $\sigma$) for three parameters. For comparison, we also plot the standard normal distribution in the orange line and a normal distribution shifted by the mean of the predictions from SBI in the blue line. In this figure, we do not see systematic biases for any of the parameters. Both of these two tests indicate that SBI predictions on fiducial cosmology are unbiased, and the discrepancy that occurs in fig.~\ref{contour} should be explained by statistical fluctuation.

Regarding the degeneracies of parameters, as depicted in fig.~\ref{contour} we calculate the correlation coefficients for each pair of three parameters. It illustrates that $\Omega_M$ strongly anti-correlate to $S_8$ ($\rho=-0.85$), inferring that choosing $S_8$ to replace $\sigma_8$ in our analysis does not alleviate the degeneracy between matter density and the standard deviation of fluctuations of dark matter field. This may be expected due to the nature of lensing statistics. It can also be noticed that the degeneracy between HAM parameter $\sigma$ and the other two cosmological parameters is weaker ($|\rho|<0.25$). It is known that the galaxy-halo connection should always be considered in cosmological analysis based on galaxy statistics, and in our case we find that considering the galaxy-halo connection does not reduce much constraining power of void lensing on cosmology.

It should be pointed out that we have not included any statistical uncertainties or systematics such as shape noise or redshift uncertainties in our analysis, and the degeneracies could become quite different after considering these uncertainties. In Appendix~\ref{appx:shape_noise} we primarily consider the effects of shape noise on our results and leave a thorough analysis for future work.

\section{Summary and Conclusion} \label{sec:sum&conclu}
Void statistics are expected to bring us more cosmological information than two point statistics. In this work, we apply the Simulation-Based Inference (SBI) method on void lensing to make constraints on cosmological parameters. Instead of building an explicit likelihood function, a neural network is constructed to learn the posterior directly from a set of simulated data. Through this method, we avoid modeling the void lensing statistics, which is quite a challenging task.

The aim of this work is to explore the potential of constraining cosmological parameters using void lensing with SBI. We show that SBI can recover a posterior distribution with unbiased mean values and slightly overestimated uncertainty. In particular, both cosmological and galaxy formation effects are incorporated into the inference pipeline, and the final posterior can at least provide unbiased predictions of cosmological parameters $\Omega_M$ and $S_8$, across different cosmologies. This is encouraging and in the future, it is planned to develop both the forward modeling process and the inference process in order to analyze the real observational data. 

In this work we have achieved these goals:

1) We use a novel method, i.e., SBI to study the accessibility and performance of void lensing on constraining cosmology. We show that by use of SBI, we achieve the Bayesian inference without an explicit likelihood and successfully recover the input model parameters. Therefore, with the help of SBI, one can overcome the problem of the absence of void lensing models, and make it possible to use void lensing to constrain cosmology. Besides, SBI also has the following two advantages: first, it does not assume a Gaussian likelihood as traditional Bayesian analysis, which may capture more information from high-order statistics or statistics containing non-Gaussian information (also the predicted likelihood should be carefully validated); second, in this work we use Neural Posterior Estimation (NPE) framework, which does not need MCMC sampling thus quite compute-efficient. We also point out that the SBI pipeline used in this work can also be easily applied to other complicated statistics that are difficult to model.

2) We construct a complete forward modeling pipeline to simulate void lensing signal beginning from cosmological parameters and discover that for void lensing statistics, two cosmological parameters $\Omega_M$ and $S_8$ parameters are degenerate, while the HAM parameter $\sigma$ does not degenerate with the other two. However, we point out that in any case both cosmology and galaxy formation process should be taken into account in the forward modeling process. In addition to cosmology, the galaxy formation process can also affect galaxy-based statistics, and incomplete modeling of galaxy-halo connections may potentially lead to a biased inference. We believe that considering the galaxy-halo connection is a crucial step in applying our method to real data.

3) We present thorough validation tests to confirm results obtained from SBI. The validation of predictions is one of the most important parts of every machine learning application literature. Multiples of validation methods have been presented in recent years and been applied in relevant papers. In this work, apart from some popular mathematical validation tests (like TARP), we also carefully validate SBI results on realizations of a single simulation, thus alleviating the influence of cosmic variance, i.e., even SBI can predict biased results on one single realization, but it gives an unbiased prediction when averaging many realizations. This validation test is more intuitive and can strongly imply the accuracy of SBI predictions. 

In order to apply our method to real data, in future works we plan to consider the following points more carefully:

1) The necessity of lightcone simulation. With the lightcone one can use the ray-tracing method to obtain a synthesis shear catalog, and one can measure the lensing signals in the same way as in real data. Furthermore, it is also convenient to consider the statistical error and systematics in shear measurements and redshift measurements. All of these reasons push us to run lightcone simulation in the next step.

2) The systematics. In modern cosmological analysis, a careful consideration of the systematics is of great importance because insufficient 
modeling of systematics will lead to biased inference of the cosmological parameters. In SBI, this is also a key point since all of the information of the posterior is from the simulated data, and if the simulated data do not contain the systematics, SBI may learn inaccurate relations between data and parameters, potentially leading to biased inferences when applied to real data. Therefore, in future work, if we want to apply SBI to real data, we must include as much known systematics in our forward modeling as possible.

3) The accuracy of fast simulation. All of the results in this work are made in a self-consistent way, i.e., the training set and the validation set are from the same forward modeling procedure, and the conclusion is that it is possible to apply SBI on void lensing to extract cosmological information. In the future, we plan to further test the accuracy of our fast simulation, i.e., if the results training on fast simulation is suitable for observables from high-fidelity simulation. Besides, we notice that \citet{NQE-cosmo} presents a hybrid training strategy that uses large fast simulation as a training set and uses small high-accuracy simulation to calibrate the training results, which can significantly reduce the computational cost and maintain high prediction accuracy. This is a valuable reference and in future work, we plan to introduce this strategy in our method and investigate the performance.

Recent developments in the observational data highlight the need for the improvement of data analysis. To maximize the extraction of information from these data, it is essential to identify statistics that preserve as much cosmological information as possible. Void statistics is one such method, as it is expected to capture information from galaxy N-point statistics. The endpoint of this way may be the field-level inference since at a fixed resolution the field includes full cosmological information. Modeling statistics that contain more information is typically more challenging, and their likelihood may not follow a Gaussian distribution. Therefore, Machine Learning (ML) techniques, such as SBI used in this work, could be helpful. Combining ML with high-order statistics could tighten constraints on the cosmological model and bring us a deeper understanding of our universe.

\begin{acknowledgements}

We acknowledge the support by the Ministry of Science and Technology of China (grant Nos. 2020SKA0110100) and National Key R\&D Program of China No. 2022YFF0503403. CS would like to thank Mingshan Xie, Linfeng Xiao for valuable discussions and suggestions on this work. CS would also like to thank ``Tree New Bee" club for supporting friendly discussion environment. HYS acknowledges the support from NSFC of China under grant 11973070, Key Research Program of Frontier Sciences, CAS, Grant No. ZDBS-LY-7013 and Program of Shanghai Academic/Technology Research Leader. CZ acknowledges the support from National Key R\&D Program of China (grant No. 2023YFA1605600). We acknowledge the support from the science research grants from the China Manned Space Project with NO. CMS-CSST-2021-A01, CMS-CSST-2021-A04. 

\end{acknowledgements}

\bibliography{sbi_VL}{}
\bibliographystyle{aa}

\begin{appendix}
\section{Influence of the shape noise}\label{appx:shape_noise}
\begin{figure}
    \centering
    \includegraphics[width=0.8\linewidth]{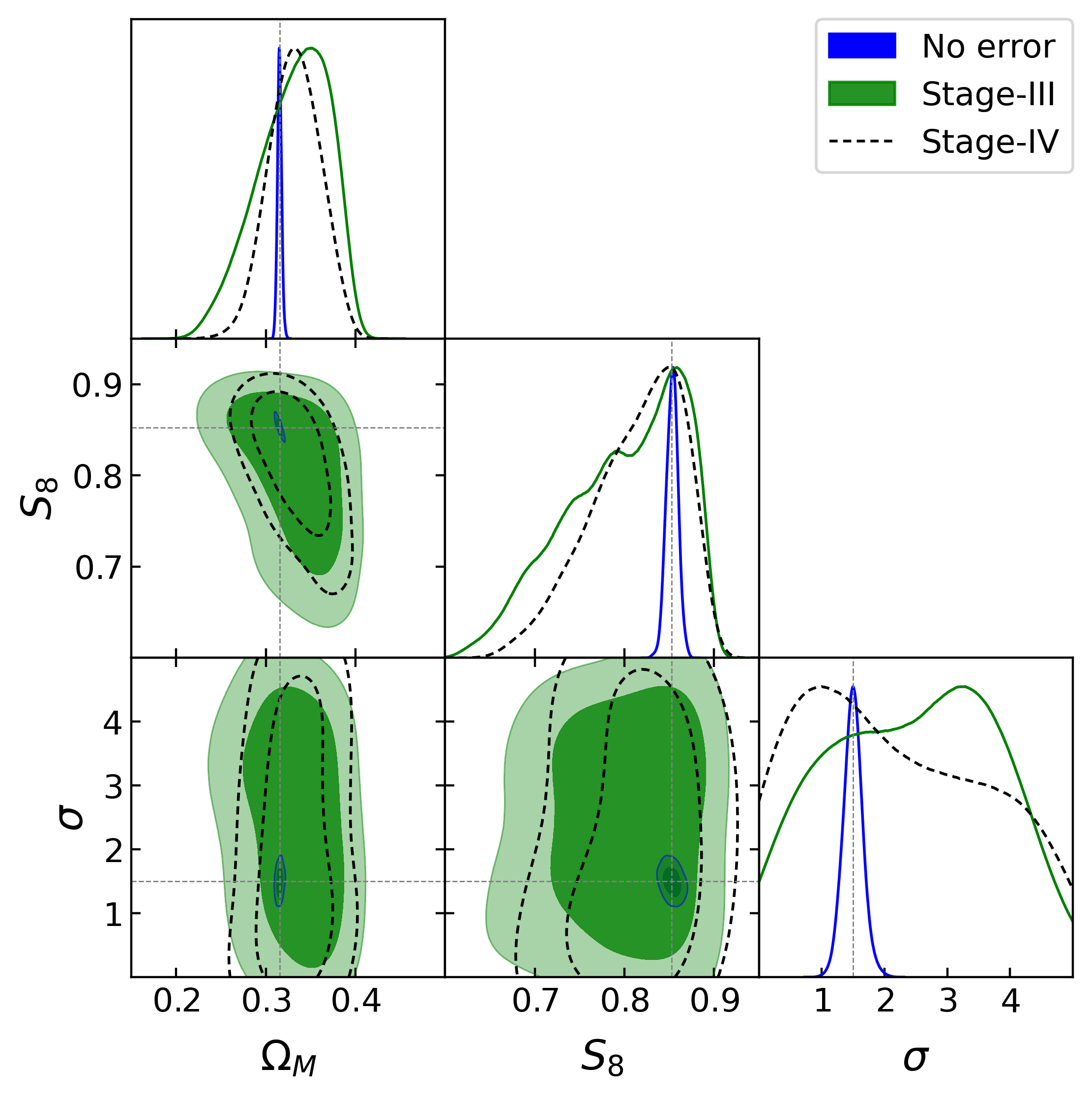}
    \caption{Constraining results after considering two kinds of shape noise, one matches Stage-III survey (shown in green filled contour) and the other matches Stage-IV (shown in black dashed contour). The noise-free case is shown in blue contour.}
    \label{cmp_w_wo_noise}
\end{figure}
The main results obtained in this work do not consider any observational uncertainties in our data vector, therefore uncertainties of our predictions are purely from cosmic variance and neural network uncertainties. One of the uncertainties in weak lensing surveys is shape noise, resulting from the stochastic alignment of background galaxies. A direct and accurate estimation of this covariance requires a ray-tracing simulation.

Due to the limitation of time and computational resources, we only run one ray-tracing simulation at fiducial cosmology to estimate the covariance and assume the covariance is independent on cosmology. We first run FastPM with the same configuration as used in generating training sets, but save the full sky lightcone from $z=1$ to $z=0$. Then we construct the discrete matter fields by mapping lightcone particles to healpix maps with $N_{\text{side}}=1024$. 40 mass maps are constructed from $z=0$ to $z=1$ with interval $\delta z=0.025$. Then we use the public ray-tracing code \textsc{dorian} to obtain the background convergence and shear fields. Then, we cut a continuous sky region with area $A\sim1000~\text{deg}^2$. Shape noise is controlled by two factors: galaxy number density $n_{\text{geff}}$ and ellipticity dispersion $\sigma_e$. For the former we choose two values $n_{\text{geff}}=6~\text{arcmin}^{-2}$ and $n_{\text{geff}}=20~\text{arcmin}^{-2}$, which match the Stage-III and Stage-IV weak lensing survey, respectively. For the latter we choose $\sigma_e=0.288$. The shape noise can be introduced by adding a noise shear $n=n_1+\mathrm{i}n_2$ to the shear $\gamma=\gamma_1+\mathrm{i}\gamma_2$ to obtain the measured ellipticity $e=e_1+\mathrm{i}e_2$: $e=(\gamma+n)/(1+\gamma n^*)$, where $n_1$ and $n_2$ are sampled from Gaussian distribution $\mathcal{N}(0,\sigma_e^2)$. Finally we can measure the void-shear cross correlation using the following estimator:
\begin{equation}
    \widehat{\Delta\Sigma}
    =\frac{\sum_{i,j}{w_{ij}e_{t,ij}\Sigma_{\text{crit},ij}}}{\sum_{ij}{w_{ij}}}, \label{esti_esd}
\end{equation}
We use the jackknife technique to estimate the covariance matrix. $n_{jk}=32$ subsamples are used in covariance estimation. Besides, for the Stage-IV survey we also consider the influence of the sky coverage by rescaling the covariance matrix with respect to the survey area.

With the covariance by hand, we can generate a set of data vectors, whose means are equal to the noise-free values and whose covariance equals what we obtained before. We then train our SBI networks using these data vectors as training sets to estimate the posteriors. The network architecture is almost the same as in analyzing noise-free data, but we adjust the learning rate to be 0.0001 and batch size to 128 to obtain a better performance and use a full-connection network to first compress data vectors to 8 dimensions in order to suppress the noise. Our results are shown in fig.~\ref{cmp_w_wo_noise}. 

It is obvious that even if we update our neural network, considering shape noise still greatly degrades the constraining power of void lensing. The posteriors of all three parameters are almost similar to the priors, especially for the HAM parameter. Reducing the shape noise level from Stage-III to Stage-IV case can increase the constraining power on cosmological parameters, but not for HAM parameters. These results, on the one hand, illustrate that our noise-free void lensing model is accurate enough (the model uncertainties shown in the blue contour are negligible compared to Stage-IV level shape noise shown in the black dashed contour), and on the other hand in order to extract more information from void lensing, data processing and neural network architecture need to be optimized. In future work, some points could be investigated: the optimized void size bins; tomography, i.e., considering combinations between multiple redshift bins; data compression, for example, using Principle Component Analysis or including embedding nets in sbi inference process, in order to reduce the dimension of the data and to suppress the influence of the statistical noise.
\end{appendix}

\end{document}